# Small anisotropy, weak thermal fluctuations, and high field superconductivity in Co-doped iron pnictide Ba(Fe$_{1-x}$Co$_x$)$_2$As$_2$

A. Yamamoto, J. Jaroszynski, C. Tarantini, L. Balicas, J. Jiang, A. Gurevich, and D.C. Larbalestier

*National High Magnetic Field Laboratory, Florida State University, Tallahassee, FL 32310, USA*

R. Jin, A.S. Sefat, M.A. McGuire, B.C. Sales, D.K. Christen, and D. Mandrus

*Materials Science & Technology Division, Oak Ridge National Laboratory, Oak Ridge, TN 37831, USA*

We performed high-field magnetotransport and magnetization measurements on a single crystal of the 122-phase iron pnictide Ba(Fe$_{1-x}$Co$_x$)$_2$As$_2$. Unlike the high-temperature superconductor cuprates and 1111-phase oxypnictides, Ba(Fe$_{1-x}$Co$_x$)$_2$As$_2$ showed practically no broadening of the resistive transitions under magnetic fields up to 45 T. We report the temperature dependencies of the upper critical field $H_{c2}$ both parallel and perpendicular to the *c*-axis, the irreversibility field $H_{irr}^c(T)$ and a rather unusual symmetric volume pinning force curve $F_p(H)$ suggestive of a strong pinning nano-structure. The anisotropy parameter $\gamma = H_{c2}^{ab}/H_{c2}^c$ deduced from the slopes of $dH_{c2}^{ab}/dT$ = 4.9T/K and $dH_{c2}^c/dT$ = 2.5T/K decreases from ~2 near $T_c$, to ~1.5 at lower temperatures, much smaller than $\gamma$ for 1111 pnictides and high-$T_c$ cuprates.

The discovery of superconductivity in iron oxypnictides[1] has attracted strong interest due to an unusual interplay of superconductivity and magnetism and extremely high upper critical fields $H_{c2}$.[2,3] Like the high temperature superconductor (HTS) cuprates, iron pnictides are layered with alternating basal Fe-As layers sandwiched between doped charge reservoir layers. Superconductivity appears upon doping a parent antiferromagnetic state with electrons[1,4] or holes,[5] resulting in high transition temperatures $T_c$ up to 55 K for the 1111 type single layer oxypnictides REFeAsO (RE = rare earth). The recently discovered 122 type AEFe$_2$As$_2$ compounds (AE = alkali or alkali earth) become superconducting with $T_c$ up to 38 K by hole doping.[6] Electron doping induced by substituting Co[7] or Ni[8] for Fe can also induce superconductivity, but, unlike the cuprates, iron-pnictides can tolerate magnetic impurities in the superconducting layers.

Our previous high-field measurements on the La, Nd and Sm 1111 oxypnictides showed very high upper critical fields $H_{c2}$ of ~65 T for LaFeAsO$_{1-x}$F$_x$ polycrystals[2] and even greater than 100 T for Nd and Sm pnictides.[3] However, the Nd and Sm 1111 compounds exhibit field-induced, thermally-activated broadening of the resistive transitions reminiscent of that in the cuprates. Concurrently, quasi-reversible magnetization was observed in LaFeAsO$_{1-x}$F$_x$ polycrystals, indicative of weak pinning of a nearly equilibrium vortex lattice.[9] It was also suggested that the higher-$T_c$ oxypnictides, like the cuprates, may have a grain-boundary weak-link problem.[9-11] On the other hand, the 122 pnictides have lower $T_c$ but also much lower anisotropy than the 1111 oxypnictides.[12-14] Here, we report on $H_{c2}(T)$ and of the irreversibility field $H_{irr}(T)$ in Ba(Fe,Co)$_2$As$_2$ single crystals. Surprisingly, we found conventional low-temperature-superconductor-like (LTS) displacements of the resistive transition in fields up to 45 T, indications for strong vortex pinning, and a very high $H_{irr}$ close to the onset of the superconducting transition at $H_{c2}$. Both $H_{c2}$ and $H_{irr}$ exceed the BCS paramagnetic limit at low $T$.

Cobalt-doped BaFe$_2$As$_2$ single crystals were synthesized by the self-flux method.[7] The present Ba(Fe$_{0.9}$Co$_{0.1}$)$_2$As$_2$ crystal has dimensions of 1.28×0.58×0.030 mm$^3$. High-field magneto-transport measurements were performed using the National High Magnetic Field Laboratory DC 45 T hybrid magnet and a 16 T Quantum Design physical property measurement system. The full critical state magnetization was measured in an Oxford 14 T vibrating sample magnetometer.

Our crystal has $T_c$ of 22 K inferred from the susceptibility measurements. Figure 1 shows magnetization hysteresis loops which exhibit a small "fish-tail" hump at 5-10 T, similar to that of (Ba,K)Fe$_2$As$_2$ (Ref. 15) and YBa$_2$Cu$_3$O$_{7-\delta}$ crystals.[16] The critical current density $J_c$ calculated from the width of the hysteresis loops using the Bean model is shown in Fig. 2. $J_c$ exhibits a rapid decrease at low-fields followed by broad maxima and a relatively weak field dependence at high-fields. The self-field $J_c = 4 \times 10^5$ A/cm$^2$ at 4.2 K is indeed high for a single crystal.

To assess mechanisms, which control the vortex pinning force $F_p = \mu_0 H J_c$, we plot in Fig. 3 the normalized pinning force $F_p/F_p^{max}$ as a function of the reduced field $h = H/H_{irr}$. Here we define the irreversibility field $H_{irr}$ at which $J_c(H)$ extrapolates to zero from the field closure of hysteretic magnetization loops. The normalized curves of $F_p(h,T)$ for $T >$ 15 K and (0.05-0.1) $< h <$ 1 collapse into a single curve described by the scaling function $h^p(1-h)^q$ with $p = 1.67$, $q = 2$ (Ref. 17). Moreover, the lower $T$, partial $F_p(h)$ curves taken at 4.2-12.5 K also exhibit the same field dependence, allowing $H_{irr}$ to be estimated down to ~10 K. The observed $F_p$ scaling, which is independent of temperature, suggests one dominant vortex pinning mechanism, while the symmetric $F_p(h)$ curves with a peak at $h$ ~0.45 imply a dense vortex pinning nano-structure, perhaps due to an inhomogeneous distribution of cobalt ions, which produces a locally varying order parameter. This scenario is consistent with a spatial variation of K in (Ba,K)Fe$_2$As$_2$ single crystals [18] and similar to the field-induced pinning by oxygen deficient regions in YBa$_2$Cu$_3$O$_{7-\delta}$.[16]

The results of magneto-transport measurements in fields parallel to the *c*-axis up to 45 T are shown in Fig. 4. It is clear that the *R*(*T*) curves are displaced to lower temperatures upon increasing fields, but also that they do not noticeably broaden. The transition widths $\Delta T$ defined by the 90% and 10% points on *R*(*T*) do not exceed 2-3 K. The transitions with fields parallel to the *ab*-plane were similarly sharp up to 45 T. This lack of broadening of the resistive transitions under field is in strong contrast to some of the 1111 oxypnictides[2,3,19] and rather similar to a conventional LTS like Nb$_3$Sn.[20]

The combined high field magneto transport and magnetization analyses enable us to obtain the magnetic phase diagram. The temperature-dependent resistive $H_{c2}(T)$ defined by 90% of $R_n$ is shown in Fig. 5. Both $H_{c2}^{ab}$ and $H_{c2}^{c}$ exhibit almost linear temperature dependence near $T_c$ with slopes of $dH_{c2}^{ab}/dT = 4.9$ T/K and $dH_{c2}^{c}/dT = 2.5$ T/K. The anisotropy $\gamma = H_{c2}^{ab}/H_{c2}^{c}$ varies from ~ 1.5 to 2 as *T* increases (see inset of Fig. 5). This temperature-dependent $\gamma$ is consistent with multi-band superconductivity, however $\gamma = 1.5$-2 is significantly lower than $\gamma = 5$-10 measured on the 1111 oxypnictides.[3,21,22] The low temperature $H_{c2}$ extrapolates to > 60 T, much larger than the Werthamer-Helfand-Hohenberg extrapolation $H_{c2}^{c}(0) \sim 0.69T_c|dH_{c2}/dT|_{Tc} \sim 38$ T, indicating unconventional $H_{c2}(T)$ behavior. Moreover, even at $T = T_c/2$, the observed $H_{c2}^{ab}$ already exceeds the BCS paramagnetic limit $H_p = 1.84T_c \sim 40.5$ T. Extrapolations of the $H_{c2}(T)$ data in Fig. 5 suggest $H_{c2}^{ab}(0) \sim 70$ T and $H_{c2}^{c}(0) \sim 50$ T, comparable to estimates of ~70 T for (Ba,K)Fe$_2$As$_2$[12-14] and ~65 T for the optimally doped LaFeAsO$_{0.89}$F$_{0.11}$,[2,19] though much smaller than $H_{c2}(0) > 100$ T for Nd and Sm oxypnictides.[3]

Now we discuss the relationship of $H_{c2}$ and $H_{irr}$ defined conventionally as the field at which $J_c(H)$ extrapolates to zero, which gives $H_{irr}(T)$ close to the onset of the flux flow resistance at $R(T) = 0.1R_n$. The lack of the field-induced broadening of *R*(*T*) and the fact that $H_{c2}$ and $H_{irr}$ are not very different indicate the LTS-like behavior, unlike HTS cuprates in which $H_{irr}$ is well below $H_{c2}$ due to strong thermal fluctuations of vortices. This LTS-like behavior is also consistent with weak thermal fluctuations, as follows from the estimation of the Ginzburg number[23] $Gi = (2\pi k_B T_c \mu_0 \lambda_0^2/\Phi_0^2 \xi_c)^2/2 \sim 6.8 \times 10^{-5}$, much lower than for the least anisotropic cuprate YBa$_2$Cu$_3$O$_{7-\delta}$ (~$10^{-2}$) and polycrystalline NdFeAs(O,F) and SmFeAsO$_{1-\delta}$ (~$10^{-2}$) and LaFeAs(O,F) (~$3.4 \times 10^{-4}$),[3] and even for clean MgB$_2$ (~$2.0 \times 10^{-4}$). Here the London penetration depths $\lambda_{ab} = 160$ nm and $\lambda_c \sim \gamma \lambda_{ab} = 320$ nm were estimated from the lower critical field $H_{c1}^{ab} \sim \Phi_0(\ln \kappa + 0.5)/4\pi \lambda_{ab} \lambda_c$ measured from the deviation of the diamagnetic magnetization using a SQUID magnetometer, where $\kappa \sim 65$ is the Ginzburg-Landau parameter. $H_{c1}(T)$ shows the usual linear temperature dependence with $dH_{c1}^{ab}/dT = 0.79$ mT/K near $T_c$, extrapolating to $H_{c1}^{ab}(0) \sim 15$ mT. The coherence lengths $\xi_{ab} = 2.44$ nm and $\xi_c \sim \gamma^{-1}\xi_{ab} = 1.22$ nm were evaluated from the $H_{c2}(0)$ extrapolations.

Given the weakness of vortex thermal fluctuations, the finite width of *R*(*T*) in our single crystal likely results from random $T_c$ inhomogeneities due to local compositional variations, proximity effect near defects, etc, which cause local fluctuations of $H_{c2}$ and the flux flow resistivity so that the curve *R*(*T*) reflects the percolative transition in a weakly inhomogeneous superconductor. In this case the onset of the superconducting transition at $H_{c2}$ represents the maximum $H_{c2}$. In turn, the onset of the global magnetic irreversibility at $H_{irr}$ may be interpreted as the field percolation threshold at which the lower $H_{c2}$ superconducting regions form an infinite percolation cluster.

In summary, our magnetization and transport measurements show very high values of $H_{c2}$ exceeding the BCS paramagnetic limit, and lack of field-induced broadening of the resistive transitions up to 45 T. The irreversibility field $H_{irr}$ is close to $H_{c2}$, indicating weak thermal fluctuations and/or strong vortex pinning. Thus, Co-doped 122 pnictide is a quasi-LTS high field superconductor with $H_{c2} >50$ T and weak anisotropy with $\gamma <2$.

We are grateful to V. Griffin, N. Craig, F. Hunte, E. E. Hellstrom and M. Putti for discussions. Work at the NHMFL was supported by the NSF Cooperative Agreement DMR-0084173, by the State of Florida, by the DOE and by AFOSR under Grant No. FA9550-06-1-0474. Work at ORNL was supported by the Division of Materials Science and Engineering, Office of Basic Energy Sciences. AY is supported by a fellowship of the JSPS.

**Figure captions**

**Fig. 1.** (Color online) Magnetization hysteresis loops at 4.2, 7.5, 10, 12.5, 15, 17.5 and 20 K. Magnetic field was applied parallel to *c*-axis.

**Fig. 2.** (Color online) Magnetic field dependence of the in-plane critical current density $J_c$. Triangles indicate broad maximum positions.

**Fig. 3.** (Color online) Normalized flux pinning force $F_p/F_p^{max}$ as a function of reduced field $h = H/H_{irr}$. Dashed line represents the fitting curve $h^{1.67}(1-h)^2$. Inset shows $F_p/F_p^{max}$ as a function of field $H$.

**Fig. 4.** (Color online) In-plane resistivity $\rho_{ab}$ as a function of temperatures under magnetic field of $\mu_0H$ = 0, 1, 3, 6, 9, 12, 15, 20, 25, 30 and 35 T. High field measurements above 15 T was performed using 45 T hybrid magnet at FSU.

**Fig. 5.** (Color online) Magnetic phase diagram of the Ba(Fe$_{0.9}$Co$_{0.1}$)$_2$As$_2$ single crystal. Filled and open symbols represent 90% and 10% of $R(T,H)$ of $R_n$, respectively. Circles and squares represent resistive $H_{c2}$ parallel to the *ab* plane and *c* axis, respectively. Upward-pointing triangles represent the irreversibility field $H_{irr}^c$ extrapolated from the pining force curves shown in Fig. 3. For $H_{irr} > 14$ T, $H_{irr}$ was estimated from the scaling of $F_p$ curves as described in text. Downward-pointing triangles mark the field of fish-tail peaks on magnetic hysteresis loops. The temperature dependencies of $H_{irr}$ and the field of the fish-tail magnetization peak can be fitted with $H^* = H^*(0)\times(1-T/T_c)^{5/4}$. Inset shows the temperature dependence of $H_{c2}$ anisotropy parameter $\gamma = H_{c2}^{ab}/H_{c2}^c$.

Figure 1

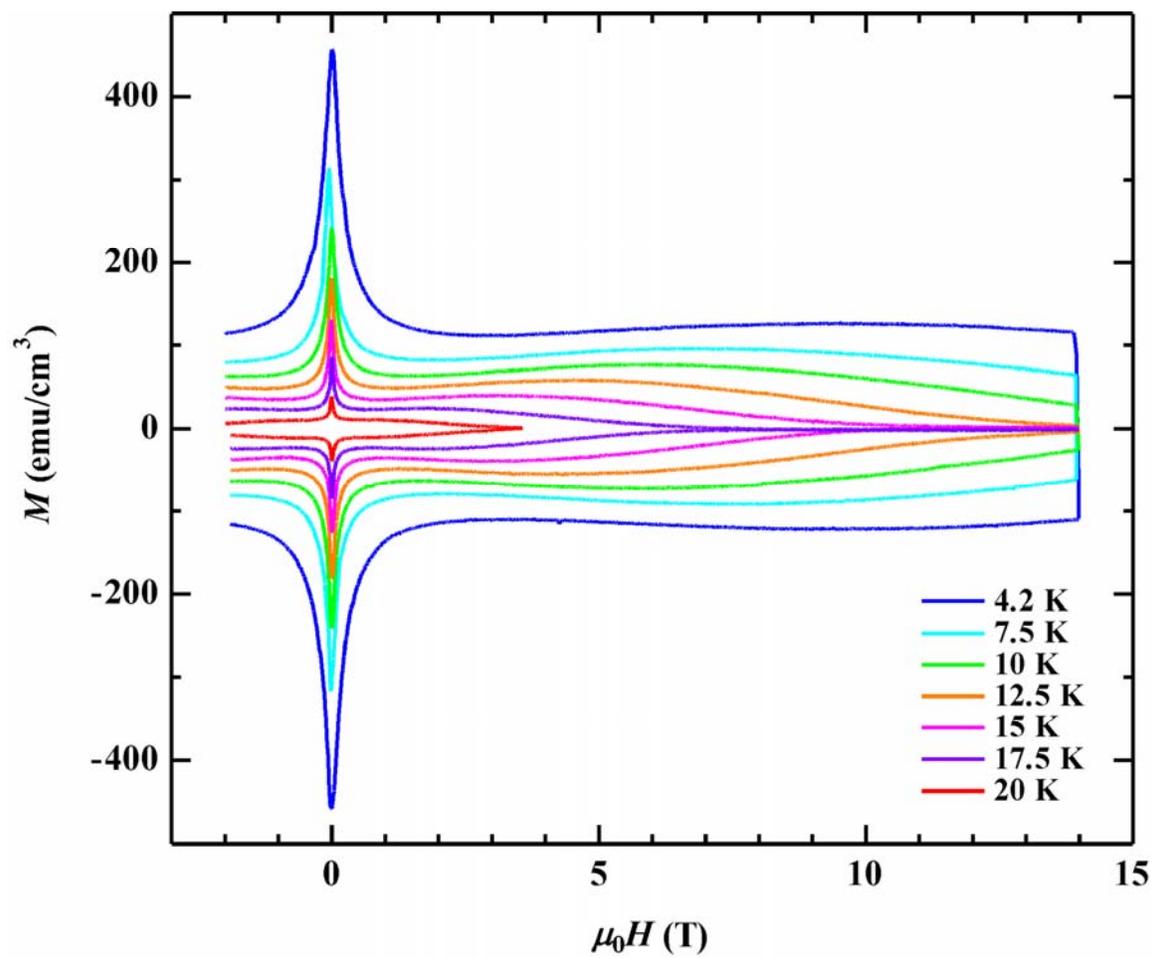

Yamamoto et al.

Figure 2

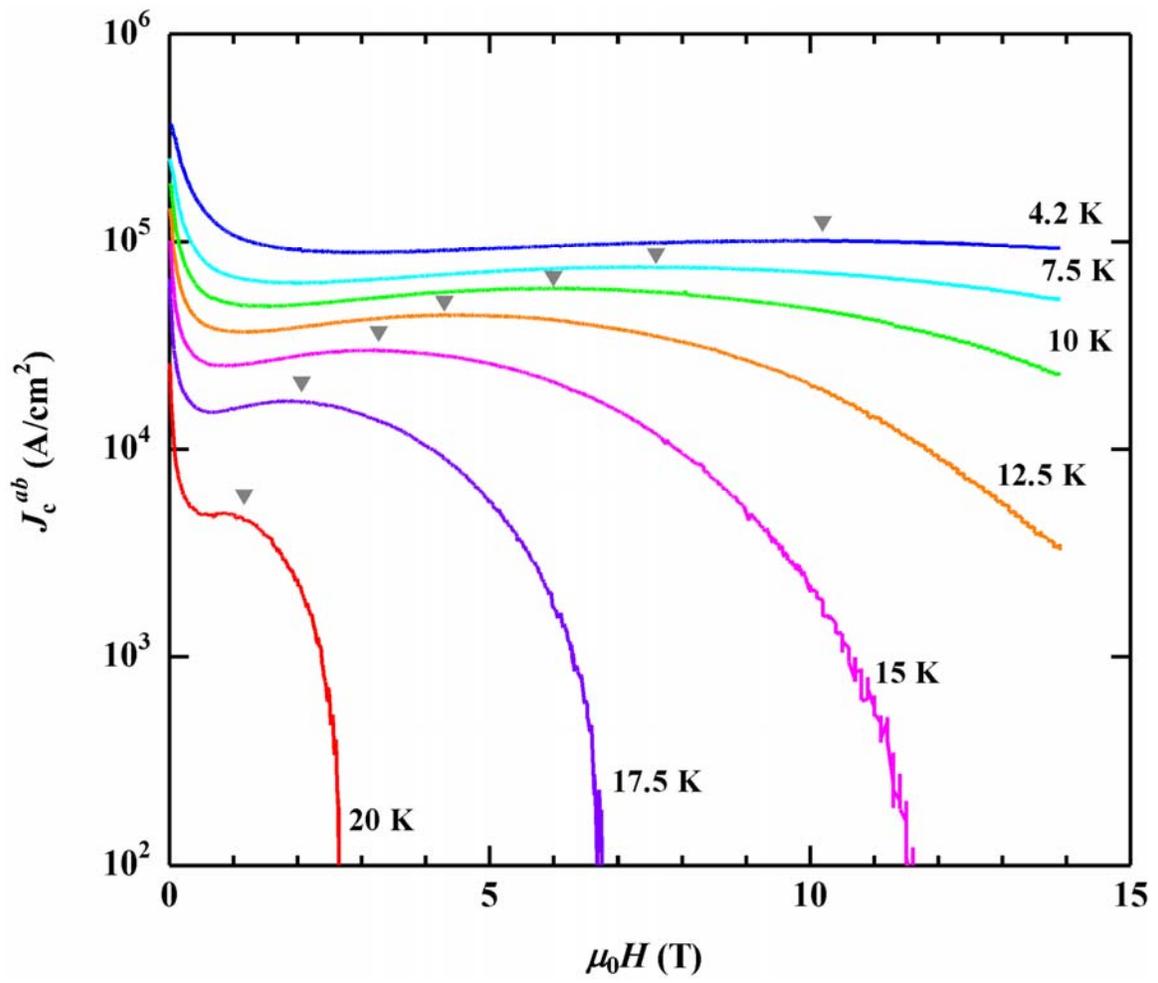

Figure 3

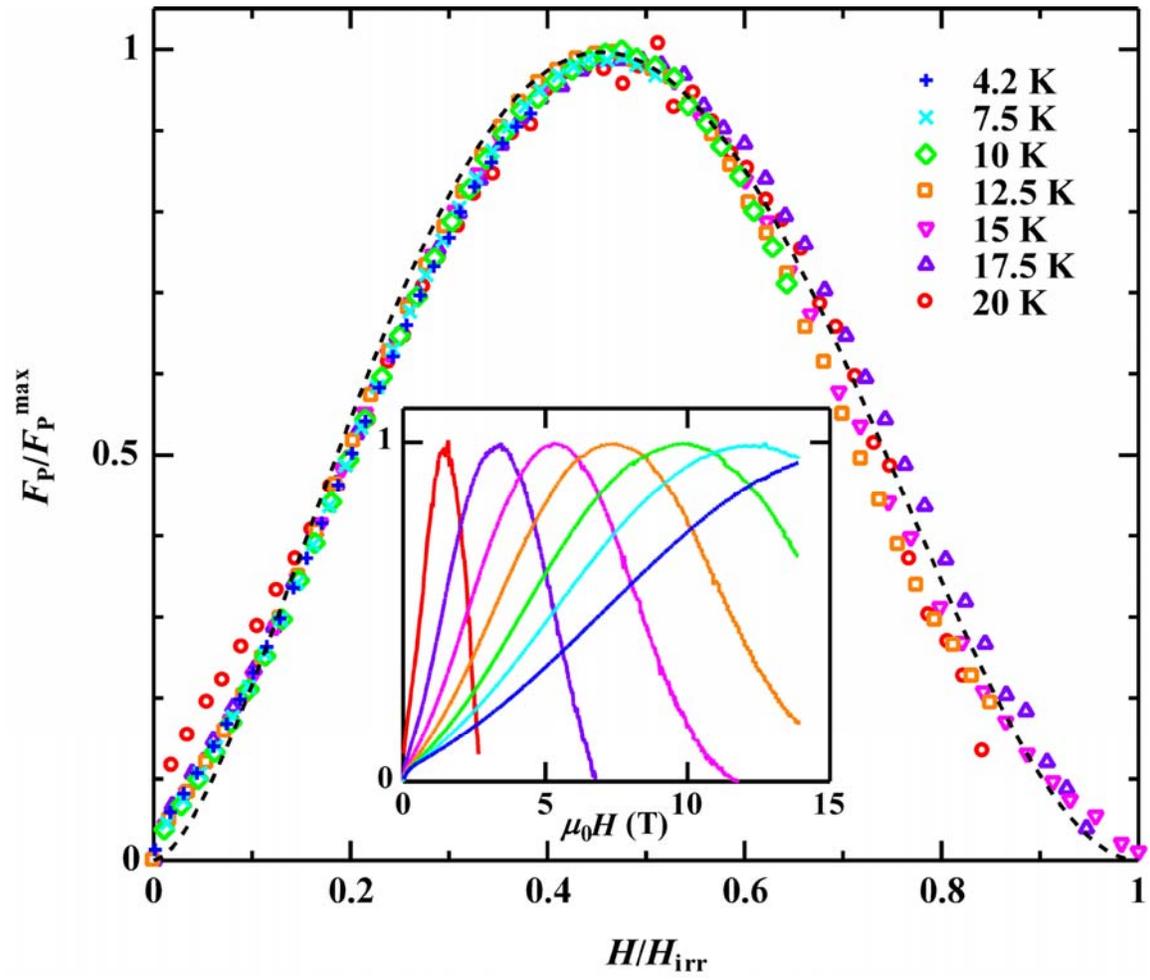

Yamamoto et al.

Figure 4

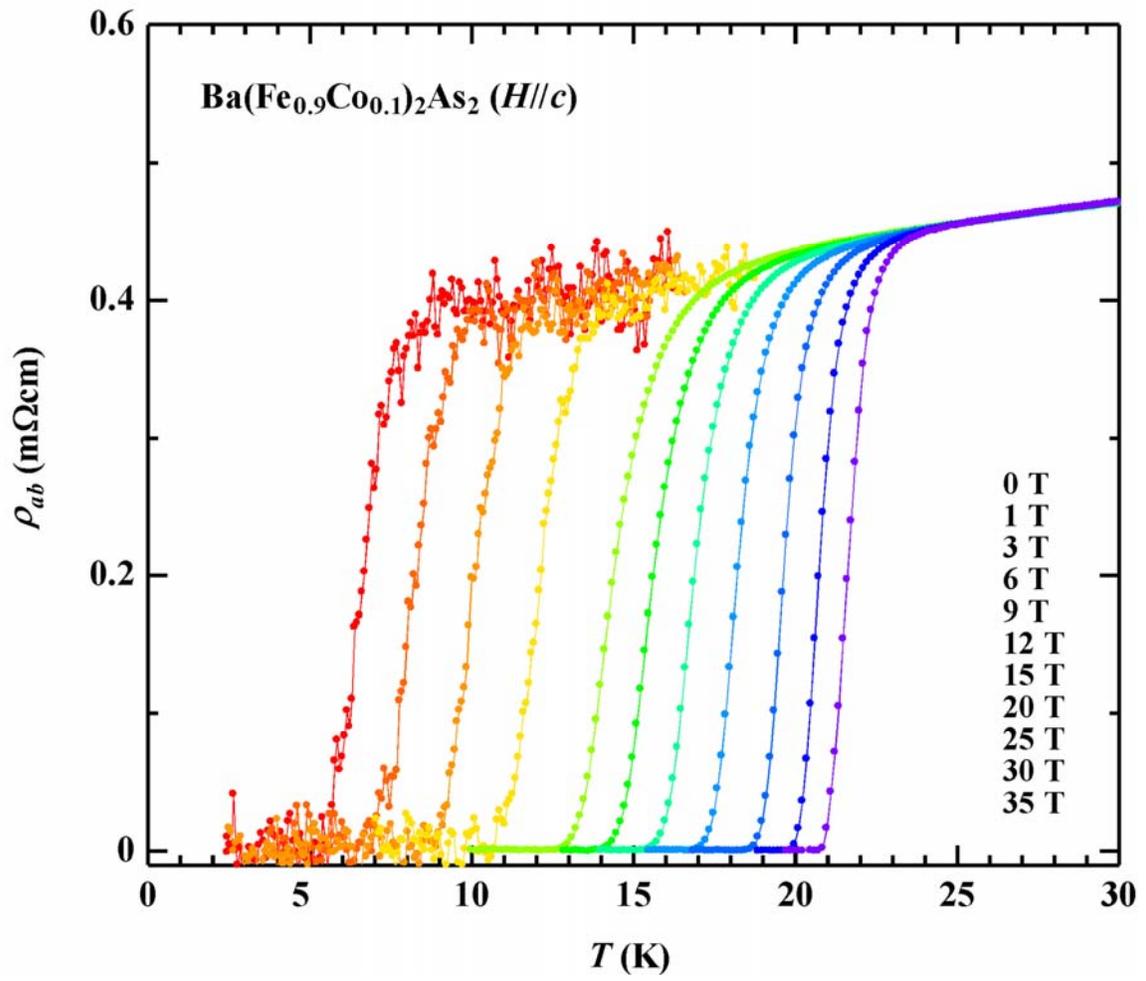

Yamamoto et al.

Figure 5

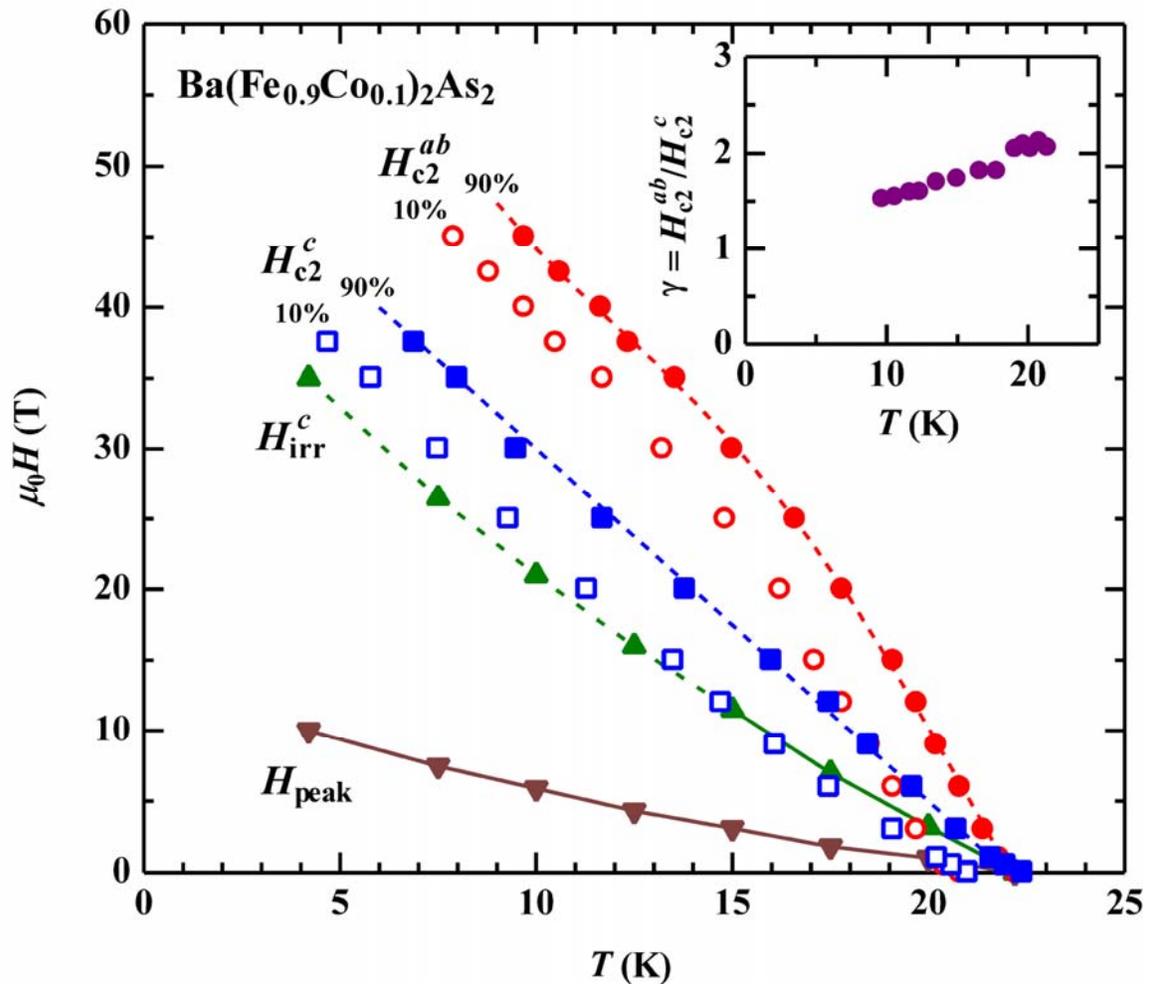

Yamamoto et al.